\documentclass[a4paper,onecolumn,fleqn]{mnras}

\usepackage{txfonts}

\usepackage[T1]{fontenc}
\usepackage{ae,aecompl}
\usepackage{graphicx}

\title[Compton scattering redistribution function ]
{On the Compton scattering redistribution function
       in plasma }

\author[J. Madej, A. R\'o\.za\'nska, A. Majczyna, M. Nale\.zyty ]{J. Madej$^{1}$
   \thanks{E-mail: jm@astrouw.edu.pl (JM) }, 
    A. R\'o\.za\'nska$^{2}$, A. Majczyna$^{3}$, M. Nale\.zyty$^{1}$   \footnotemark[1] \\
$^{1}$ Astronomical Observatory, University of Warsaw,
        Al. Ujazdowskie 4, 00-478 Warszawa, Poland          \\ 
$^{2}$ N. Copernicus Astronomical Center, Bartycka 18, 00-716 Warsaw, Poland    \\
$^{3}$ National Centre for Nuclear Research, ul. Andrzeja So\l{}tana 7, 05-400 Otwock, Poland }

\date{Accepted XXX. Received YYY; in original form ZZZ}

\pubyear{2017}

\begin{document}
\label{firstpage}
\pagerange{\pageref{firstpage}--\pageref{lastpage}}

\maketitle

\begin{abstract}
Compton scattering is the dominant opacity source in hot neutron
stars, accretion disks around black holes and hot coronae.
We collected here a set of numerical expressions of the
Compton scattering redistribution functions for unpolarized radiation
(RF) , which are more exact than the widely used Kompaneets equation. 
The principal aim of this paper is presentation of the RF by 
Guilbert (1981) which is corrected for the computational errors in the original
paper. This corrected RF was used in the series of papers on model atmosphere
computations of hot neutron stars.
We have  also organized four existing algorithms for the RF computations 
into a unified form ready to use in radiative transfer and model atmosphere codes.
The exact method by Nagirner and
Poutanen (1993) was numerically compared to all other algorithms 
in a very wide spectral range from hard X-rays to radio waves. Sample 
computations of the Compton scattering redistribution functions in thermal plasma
were done for temperatures corresponding to the atmospheres of bursting neutron stars
and hot intergalactic medium. Our formulae are also useful to the study 
Compton scattering of unpolarised microwave background radiation in hot 
intra-cluster gas and the Sunyaev-Zeldovich effect. 
We conclude, that the formulae by Guilbert (1981) and the exact quantum mechanical
formulae yield practically the same redistribution functions for gas temperatures
relevant to the atmospheres of X-ray bursting neutron stars, $T \le 10^8$~K.
\end{abstract}


\section{Introduction}

Compton scattering of unpolarized photons on free thermal electrons 
plays a crucial role in continuum and line spectrum formation in 
various astrophysical objects. The essential features of the scattering are 
a random change of direction of photon propagation and an exchange 
of energy and momentum between colliding particles.
Compton scattering is a dominant source of continuum opacity in very hot DA white dwarfs
and unmagnetized neutron stars and is responsible for the continuum 
spectrum formation in Type I X-ray bursters.
In other objects Compton scattering influences the line spectrum of OB 
giant or main sequence stars. In the X-ray domain, Compton scattering of external 
irradiation creates the Compton shoulder of fluorescent iron $K_\alpha$ 
lines at 6.4 keV in the spectra of active galactic nuclei and galactic 
black hole binaries.  

The scattering is an intrinsically strongly nonisotropic process, which also
depends on the state of the incident photon polarization.
However, herein, we consider the angle-averaged Compton scattering of 
unpolarized thermal radiation in the absence of a magnetic field. Such an
averaged process can be best described by a redistribution function (RF),
which gives the probability density of photon energy and the momentum change 
upon scattering.

Compton scattering of unpolarized radiation has been studied in a number
of papers in the literature and the most pertinent to the present study being
those of Buchler and Yueh (1976), Guilbert (1981), Nagirner and Poutanen (1993), 
Poutanen (1994), Sazonov and Sunyaev (2000) and Poutanen and Vurm (2010). 
Paper by Younsi and Wu (2013) defined general relativistic Compton 
redistribution function and its moments.

In this paper we present set of equations which define the RF derived 
by Guilbert (1981) which is corrected here for the computational errors in
the latter paper. This corrected RF was used in the series of papers on model
atmosphere computations of hot neutron stars starting from Madej (1989) and
extended also to irradiated relativistic accretion disks, cf. Madej \&
R\'o\.za\'nska (2000).

Herein we compared the procedure by Guilbert (1981) with the exact quantum
mechanical method summarized by Suleimanov et al. (2012), Appendix A, and
with two other approximate algorithms. All assume that isotropic plasma
is nondegenerate with fully relativistic electron thermal velocities. 

Furthermore, we collected  the formulae derived in other
published papers describing the redistribution of Compton scattered
photons over energies and scattering angles. Our aim was to obtain
expressions for Compton scattering cross sections and kernels (Pomraning 1973)
that would be useful in a very wide range of temperatures and frequencies.

Section 2 thus presents a list of equations and auxiliary variables that
allow for the determination of Compton scattering cross-sections in unified
form following various methods. We do not aim to discuss or evaluate 
the corresponding physical assumptions or approximations used in the original papers.
Instead, our paper is rather a description and purely numerical tests of
the new code (now publicly available) for RF computations using several
available algorithms.

Section 3 presents the exact Compton redistribution function derived by 
Nagirner and Poutanen (1993), Poutanen and Vurm (2010) and Suleimanov et al. (2012). 
Section 4 presents the Compton redistribution formulae by Guilbert
(1981), but corrected for computational errors in the original paper. 
Guilbert's and exact approaches implement Klein-Nishina scattering cross-sections
from electrons at rest. For a completeness, both angle-dependent cross-sections
were compared to a third, approximate formula obtained assuming that electron
scattering is isotropic with classical Thomson cross-sections in the electron
rest frame (Poutanen \& Svensson 1996; Suleimanov et al. 2012). Fourth formula
was taken from Sazonov and Sunyaev (2000), see Eqs. 7a-7d therein.

\section{Compton scattering redistribution function}

Here, the key variable is $R\, (\nu, \nu^\prime, \eta)$, which denotes the probability of
scattering a photon with an initial frequency $\nu$ at a unit solid angle $d\Omega$
and unit frequency range at a final frequency $\nu^\prime$ (in Hz), counted
per unit distance along the ray path. Variable $\eta=\cos\theta$ is the cosine
of a scattering angle $\theta$. Variable $R$ is equal to the differential scattering
coefficient $\sigma (\nu \rightarrow \nu {^\prime}, \vec n \cdot \vec n {^\prime})$
defined by Pomraning (1973), see Eq. 1-31 therein.

Function $R\, (\nu, \nu^\prime, \eta)$ was also integrated over the solid angle $d\Omega$. 
The angle-integrated redistribution function $R(\nu ,\nu^\prime )$ (the Compton
scattering kernel) is then given in Hz$^{-1}$.

The scattering electrons in the plasma of temperature $T$ have a thermal relativistic
Maxwellian velocity distribution given by
\begin{equation}
f_e\,(p) = {1\over {4\pi\, \Theta\, K_2(1/\Theta)}}\, \exp(-\gamma/\Theta) \, ,
   \hskip10mm {\rm where}\hskip3mm \Theta = kT / m_e c^2  \, .
\label{equ:int1}
\end{equation}
and $\gamma$ denotes the electron Lorentz factor.

The following sections 3-5 present four different algorithms for the computation
of the Compton scattering redistribution function in a unified form, suitable for
the radiative transfer calculations. Algorithms were either corrected for 
fatal algebraic errors (Guilbert 1981) or reexpressed to a more optimal form
than that given in Suleimanov (2012). We apply the original symbols and variables
used in those papers where it was useful.

Photon energies below can be expressed in units of the electron rest mass
\begin{equation}
\epsilon = h\nu /m_e c^2 \hskip9mm {\rm and} \hskip9mm 
   \epsilon_1 = h\nu^\prime /m_e c^2 \, .
\end{equation}
Note, that the variable $x$ denotes dimensionless temperature in the
following section \ref{sec:sec3}; whereas the same symbols $x$ and $x_1$
denote the photon energies $\epsilon$ and $\epsilon_1$ in sections
\ref{sec:sec4} and \ref{sec:sec8}.


\section{Exact quantum mechanical formula }
\label{sec:sec4}

The redistribution function for Compton scattering has been derived from fully 
relativistic calculations (Nagirner and Poutanen 1993; Poutanen and Svensson 1996; 
Poutanen and Vurm 2010; Suleimanov et al. 2012).
Here, the probability of scattering a photon of dimensionless energy $x_1$ to energy
$x$ with the cosine of a scattering angle $\eta=\cos\theta$, equals:

\begin{equation}
R(x,x_1,\eta) = {3\over 8} \int \limits_{\gamma_*}^\infty f_e\,(p)\,
  R(x,x_1,\eta,\gamma)\,d\gamma = {3\over{32\pi\,\Theta\, K_2(1/\Theta) }}
  \int \limits_{\gamma_*}^\infty R(x,x_1,\eta,\gamma) \,
  \exp(-\gamma/\Theta)d\gamma \, ,
\label{equ:int2}
\end{equation}
where
\begin{eqnarray}
\gamma_* (x,x_1,\eta) &=& \left( x-x_1+Q\sqrt{1+2/q}\, \right) /2 \, , \\
  Q^2 &=& (x-x_1)^2 + 2q \, , \hskip10mm q= x x_1 (1-\eta) \, .
\label{equ:int3}
\end{eqnarray}
Setting a new variable $u=(\gamma-{\gamma_*})/\Theta$, then $du=d\gamma/\Theta$,
and thus we obtain
\begin{equation}
R(x,x_1,\eta) = {3 \over{32\pi\,K_2(1/\Theta)}}\,\exp(-{\gamma_*}/\Theta)
  \int \limits_0^\infty R(x,x_1,\eta,u\Theta+{\gamma_*})\,\exp(-u)\, du\, .
\label{equ:int4}
\end{equation}
The above integral can also be calculated with the Gauss-Laguerre quadrature.

\subsection{Calculating the integrand}

The kernel of the redistribution function in Eqs.~\ref{equ:int2} \& \ref{equ:int4}
is exactly given by the analytical expression (Aharonian \& Atoyan 1981; 
Nagirner \& Poutanen 1994; Suleimanov et al. 2012)
\begin{equation}
R(x,x_1,\eta,\gamma) = {2\over Q} + {{q^2-2q-2} \over{q^2 }} \,
  \left( {1\over{a_-}} - {1\over{a_+}} \right) + {1\over {q^2}}
  \left( {{d_-}\over{a^3_-}} + {{d_+}\over{a^3_+}} \right) \, ,
\label{equ:int6}
\end{equation}
where
\begin{eqnarray}
 a^2_- &=& \, (\gamma-x)^2 + {{1+\eta}\over{1-\eta}} \, , \hskip10mm
 a^2_+ = \, (\gamma+x_1)^2 + {{1+\eta}\over{1-\eta}} \, , \\
 d_\pm &=& \, (a^2_+ - a^2_- \pm Q^2 ) / 2 \, ,  \\
 Q^2 &=& (x-x_1)^2 + 2q \, , \hskip10mm q= x x_1 (1-\eta) \, .
\label{equ:int7}
\end{eqnarray}

Unfortunately, the direct use of Eq.~\ref{equ:int6} is not possible in 
some numerical applications, both at the long wavelength part of an X-ray
burst spectra and for tracing the scattering of relic radiation in 
galaxy clusters. This is due to a catastrophic cancellation of significant
digits in the floating point representation of the last term in Eq.~\ref{equ:int6}.

\subsection{Extreme temperature differences}

Consider the Compton scattering of soft (i.e. cold) photons in a hot cloud of electrons,
when $x \ll 1$. Since variable $\gamma$ equals or exceeds 1, then the values
of variables $a_-$ and $a_+$ approach each other extremely closely. Therefore, 
difference of powers $a_-^\alpha - a_+^\alpha$ is inaccurately computed 
when all the bits representing both numbers in the computer processor compensate
each other, also in the double precision calculations. Note that noise in
the numerical
values of the above difference is amplified by the factor $1/q^2$,
sometimes rising quite arbitrarily above $10^{30}$ or even much more.
Consequently, Eq.~\ref{equ:int10} for function $R(x,x_1,\eta,\gamma)$
yields meaningless results due to the catastrophic cancellations.

The problem of cancellation of terms in some regions of the parameters space
was early recognized by Kershaw et al. (1986). Solution of the cancellation
problem was also proposed by Nagirner and Poutanen (1993), section 7, and
Poutanen and Vurm (2010), appendix E. In this paper solution of the cancellation
was obtained by manipulation of the Eq.~\ref{equ:int6}.

After a algebraic calculations Eq.~\ref{equ:int6} was transformed
into the form in which the cancellation problem does not exist
\begin{equation}
R(x,x_1,\eta,\gamma) = {2\over Q} + \left[ { (a^2_- +a_- a_+ +a^2_+) Q^2 
-(x^2 -2\gamma\, (x+x_1) - x_1^2)^2 - a_+ a_- (a_- - a_+)^2 \over{2q^2 a^3_+ a^3_-}}
- {{q-2}\over {q a_+ a_-}} \right] \left( a_- - a_+ \right) \, ,
\label{equ:int10}
\end{equation}
\noindent where one must substitute
\begin{equation}
a_- - a_+ = {{a^2_- -a^2_+} \over {a_- + a_+}} =
{{x^2 -2\gamma \, (x + x_1) - x_1^2} \over {a_- + a_+}} \, . \hspace{3cm}
\label{equ:int11}
\end{equation}
Eqs.~\ref{equ:int10}-\ref{equ:int11} are numerically fully useful and are analytically 
identical with Eq.~\ref{equ:int6}.
\vspace{2mm}

Note, that the denotation of photon energies with and without the subscripts, $x$
(final energy) and $x_1$ (initial energy) was reversed in source papers and,
therefore, in this section as compared to Guilbert (1981), see section \ref{sec:sec3}. 

Here, we have arbitrarily chosen the substitution $x=\epsilon$ and $x_1=\epsilon_1$
for the initial and final photon energies, respectively. Then, the exact Compton
scattering redistribution function is given by (procedure 1),
\begin{equation}
  R_1 (\nu, \nu^\prime, \eta)={2\pi} \times {\epsilon \over \nu}
  \times {\epsilon_1 \over \epsilon } \times R(x,x_1,\eta)
  \times \exp\left( {{\epsilon - \epsilon_1}\over \Theta}\right) \hspace{1.5cm}
  {\rm in}\,\, {\rm Hz}^{-1}
\label{equ:int12}
\end{equation}
following the symmetry and rescaling properties of the function $R(x,x_1,\eta)$.
See Pomraning (1973) and Nagirner \& Poutanen (1994), for example.

\section{{\bf Redistribution function by } Guilbert (1981) }
\label{sec:sec3}

Guilbert (1981) folded the Klein-Nishina scattering cross section with
the relativistic Maxwellian velocity distribution (see Eq. (1) in section 2). 

The probability density of scattering a photon of energy $\epsilon$ to $(s,s+ds)$
is then
\begin{equation}
P(\epsilon, s, \theta, x) = -{3\over{64\pi^2}}\,{1\over{Es^2}}\, 
  {x\over{K_2(x)}}  \int \limits_{\gamma_{\rm min}}^\infty 
  F(\epsilon, \epsilon_1,\theta,\gamma) \,\exp(-\gamma/\Theta)\, d\gamma \, , \\
\label{equ:Guilb}
\end{equation}
where the photon energy $\epsilon_1$ after scattering is expressed by the inverted 
variable $s=\epsilon / \epsilon_1$. 

The inverted dimensionless gas temperature $x = m_e c^2 / {kT} = 1/\Theta$
and $K_2(x)$ is the modified Bessel function. Other auxiliary variables are 
defined by
\begin{eqnarray}
A &=& 1-s\, ,\hskip10mm B=\epsilon(1-\cos\theta)\, , \hskip10mm E = (1-2s\cos\theta+s^2)\, ^{1/2}\, ,\\
\gamma_{\rm min} &=& \left\{ \left[ {E^2 \over{E^2 + B^2}} \left( 1-
  {A^2 \over {E^2 +B^2}} \right) \right]^{1/2} - {AB \over{E^2 +B^2}} \right\}^{-1} \, .
\label{equ:int2}
\end{eqnarray}
Changing the variables in the integral yields 
\begin{eqnarray}
P(\epsilon, s, \theta, x) &=& -{3\over{64\pi^2}}\,{1\over{Es^2}}\, 
  {x\over{K_2(x)}} \, {\exp (-x\, \gamma_{\rm min})\over x}
\int \limits_0^\infty 
  F(\epsilon, \epsilon_1,\theta, t/x +\gamma_{\rm min}) \, \exp(-t)\, dt\, ,
\label{equ:int3}
\end{eqnarray}
The integral can then be numerically calculated using the Gauss-Laguerre quadrature.
Computing the integrand  $ F(\epsilon, \epsilon_1,\theta,\gamma) $ is described
in detail in Appendix A.

A further change of the variables $s\rightarrow \epsilon_1$ yields
\begin{eqnarray}
P(\epsilon,\epsilon_1,\theta,x) &=& P(\epsilon,s,\theta,x)\frac{ds}{d\epsilon_1}
  = -\frac{\epsilon}{\epsilon_1^2} P(\epsilon, s, \theta, x) \, ,  \\
P(\epsilon,\epsilon_1,\theta,x) &=& {3\over{64\pi^2}}\,{1\over{E\epsilon}}\, 
  {x\over{K_2(x)}} \, {\exp (-x\, \gamma_{\rm min})\over x}
\int \limits_0^\infty 
  F(\epsilon, \epsilon_1,\theta, t/x +\gamma_{\rm min}) \, \exp(-t)\, dt\, .
\label{equ:int4}
\end{eqnarray}

\noindent
Finally, the resulting Compton redistribution function obtained via this method (procedure 2)
is given as
\begin{equation}
 R_2 (\nu,\nu^\prime, \eta)={2\pi} \times {\epsilon \over \nu} \times 
  P(\epsilon,\epsilon_1, \theta, x) \hspace{2.25cm} {\rm in}\,\, {\rm Hz}^{-1}
\label{equ:int0}
\end{equation} 
$P$ is the probability of scattering for unit interval of energy $\epsilon$, factor
$\epsilon/\nu = \epsilon_1/\nu^\prime = h/(m_e c^2)$ changes to probability for
1 Hz interval and the factor $2\pi$ results from integration over azimuth.

\section{Other approximate formulae }
\label{sec:sec8}

\subsection{Arutyunyan and Nikogosyan (1980) }

The third (and the earliest) method of computing the differential Compton scattering
cross section follows from the approximation by Arutyunyan and Nikogosyan (1980);
see also Poutanen and Svensson (1996) and Suleimanov et al. (2012)
\begin{equation}
R(x,x_1,\eta) = {1\over {8\pi Q}} \, {{\exp(-\gamma_* /\Theta)} \over{K_2 (1/\Theta) }} 
\label{equ:int13}
\end{equation}
where $\gamma_* (x,x_1,\eta)$ was defined in Eq.~\ref{equ:int3}.
Consequently, the approximate Compton redistribution function is given by (procedure 3),
\begin{equation}
  R_3 (\nu, \nu^\prime, \eta)={2\pi} \times {\epsilon \over \nu}
  \times {\epsilon_1 \over \epsilon } \times R(x,x_1,\eta)
  \times \exp\left( {{\epsilon - \epsilon_1}\over \Theta}\right) \hspace{1.5cm} {\rm in}\,\,
 {\rm Hz}^{-1}
\end{equation}


\subsection{Sazonov and Sunyaev (2000) }

Sazonov and Sunyaev (2000) derived the approximate Compton redistribution function 
for monochromatic radiation of $h\nu \le 50 $ keV, which is valid in partly relativistic 
thermal plasma, $kT_e \le 25 $ keV. Eqs. 7a-d of their paper can be rewritten as:
\begin{eqnarray}
R_4 (\nu,\nu^\prime, \eta) &=& {2\pi \over \nu} \times {3\over {32\pi}} \sqrt{2\over{\pi\Theta}}\,
 \,\, {\epsilon_1 \over {(\epsilon^2-2\,\epsilon\, \epsilon_1\eta+\epsilon_1^2)^{1/2} }}
 \left\{1+\eta^2+\left( {1\over 8} -\eta -{63 \over 8}\eta^2
 + 5\eta^3 \right)\Theta - {\eta\, (1+\eta)\over 2} S^2     \right.  \\
 && \left. - {3\, (1+\eta^2)\over {32\, (1-\eta)^2 }} {S^4 \over\Theta} 
 + \eta \, (1-\eta^2) S \epsilon + {1+\eta^2 \over {8\, (1-\eta) }} {S^3 \over\Theta} \, \epsilon
 + (1-\eta)^2 \, \epsilon \, \epsilon_1 \right\} \exp \left[ -{S^2 \over {4\, (1-\eta) \Theta }} \right]
 \hspace{1.0cm} {\rm in}\,\, {\rm Hz}^{-1}   \nonumber
\label{equ:k41}
\end{eqnarray}
where 
\begin{equation}
 S = {2^{1/2}\, (1-\eta)^{1/2} \over {(\epsilon^2 - 2\,\epsilon\, \epsilon_1\eta+\epsilon_1^2)^{1/2} }}
 \, \left[ \epsilon_1 -\epsilon + \epsilon\, \epsilon_1  \, (1-\eta) \right] \, .
\label{equ:k42}
\end{equation}

\section{Compton scattering coefficient}

All the above redistribution functions, $R_1$ to $R_4$, depend on the  cosine 
of the scattering angle $\eta$, but have been integrated already over the
azimuth $\phi$. The total probability of scattering a photon from frequency
$\nu$ to $\nu^\prime$ at any angle is then given by the integral
\begin{equation}
\label{equ:int18}
R_i (\nu,\nu_1) = \int \limits_{-1}^{+1} R_i (\nu,\nu_1,\eta)\, d\eta\,, \hspace{0.6cm}
  i=1,\ldots,4 \hspace{2.0cm}  {\rm in \,\,  Hz}^{-1}
\end{equation}
That integral is equivalent to the Legendre moment of the zeroth order of the
angle-dependent scattering probability (Pomraning 1973, p. 191).

Functions $R_i (\nu,\nu_1)$ of Eq.~\ref{equ:int18} were computed here numerically
using trapezoidal rule, where the interval of integration [-1,+1] was divided into
$2\times 10^3$ or more equal parts. The integrand $R_i (\nu,\nu_1,\eta)$ was
computed with the standard 15-point Gauss-Laguerre quadrature.

The frequency-dependent Compton scattering coefficient $\sigma_\nu$ is simply related
to the total probability $P(\nu)$ of scattering (Guilbert 1981) 
\begin{equation}
\sigma_{\nu, i} \, = \sigma_{\displaystyle \scriptscriptstyle T}
\int \limits _{0}^{\infty} R_i (\nu, \nu^\prime) \, d\nu ^\prime \, ,
\hspace{0.6cm} i=1,\ldots,4  \hspace{2.0cm} {\rm in  \,\, cm}^{2}
\end{equation}
where $\sigma_{\displaystyle \scriptscriptstyle T} = 6.65\times 10^{-25}$ cm$^{2}$
is the classical Thomson cross-section.

    \begin{figure}
    \resizebox{0.7\hsize}{!}{\rotatebox{0}{\includegraphics{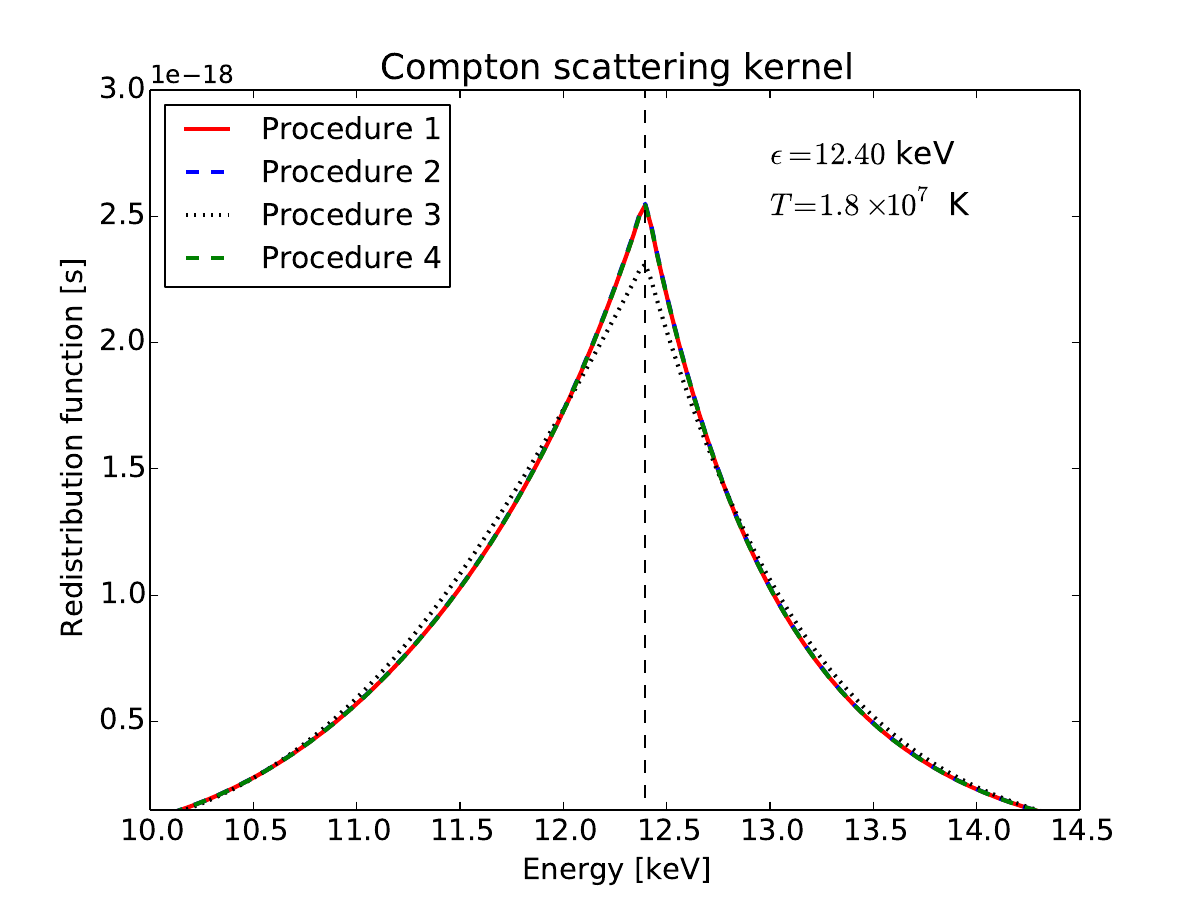}}}
    \vspace{0mm}
    \caption[]{Angle-integrated Compton scattering redistribution functions for X-ray photons of
initial wavelength $\lambda = 1 \AA$ (initial energy $\epsilon = 12.4$ keV) in gas of electron
temperatures $T=1.8\times 10^7$K. Note, that the formulae by Guilbert (1981) predict practically the
same Compton redistribution function as the exact quantum-mechanical formulae by Suleimanov et al. (2012),
compare the solid red line and dashed blue line. The approximate formulae by Arutyunyan and Nikogosyan
(1980) yield a slightly different function, see the black dotted line. Green dashed line results 
from RF by Sazonov and Sunyaev (2000) and matches the exact RF.}
    \label{fig:fig1}
    \end{figure}
    
    \begin{figure}
    \resizebox{0.7\hsize}{!}{\rotatebox{0}{\includegraphics{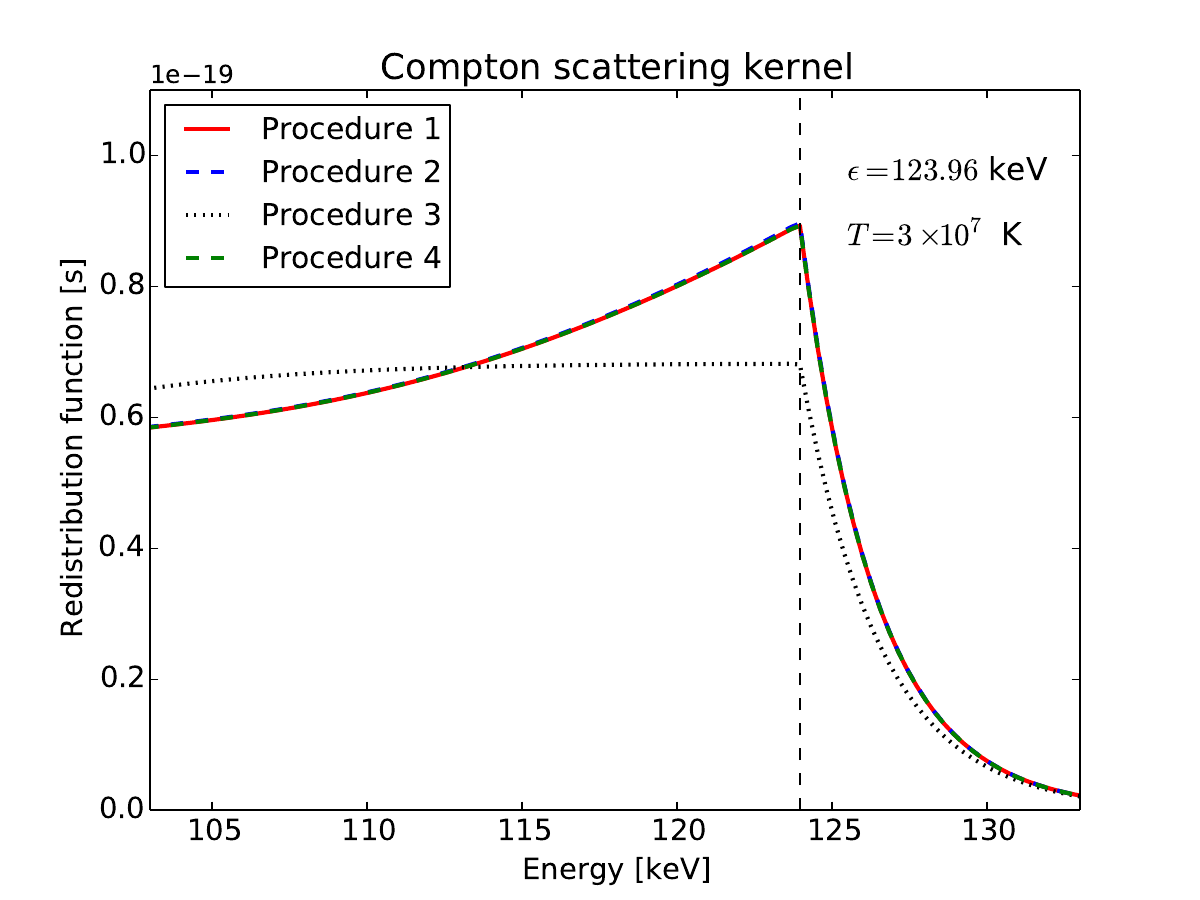}}}
    \vspace{0mm}
    \caption[]{Angle-integrated Compton scattering redistribution functions for X-ray photons of
initial wavelength $\lambda = 0.1 \AA$ (initial energy $\epsilon = 124$ keV) in a gas
of electron temperature $T=3\times 10^7$K. Again, the formulae by Guilbert (1981) and
Sazonov \& Sunyaev (2000) yield practically the
same Compton RF as the exact quantum-mechanical formulae by Suleimanov et al. (2012). 
The black dotted line denote the approximate RF obtained from Arutyunyan and Nikogosyan (1980). }
    \label{fig:fig2}
    \end{figure}
    
    \begin{figure}
    \resizebox{0.7\hsize}{!}{\rotatebox{0}{\includegraphics{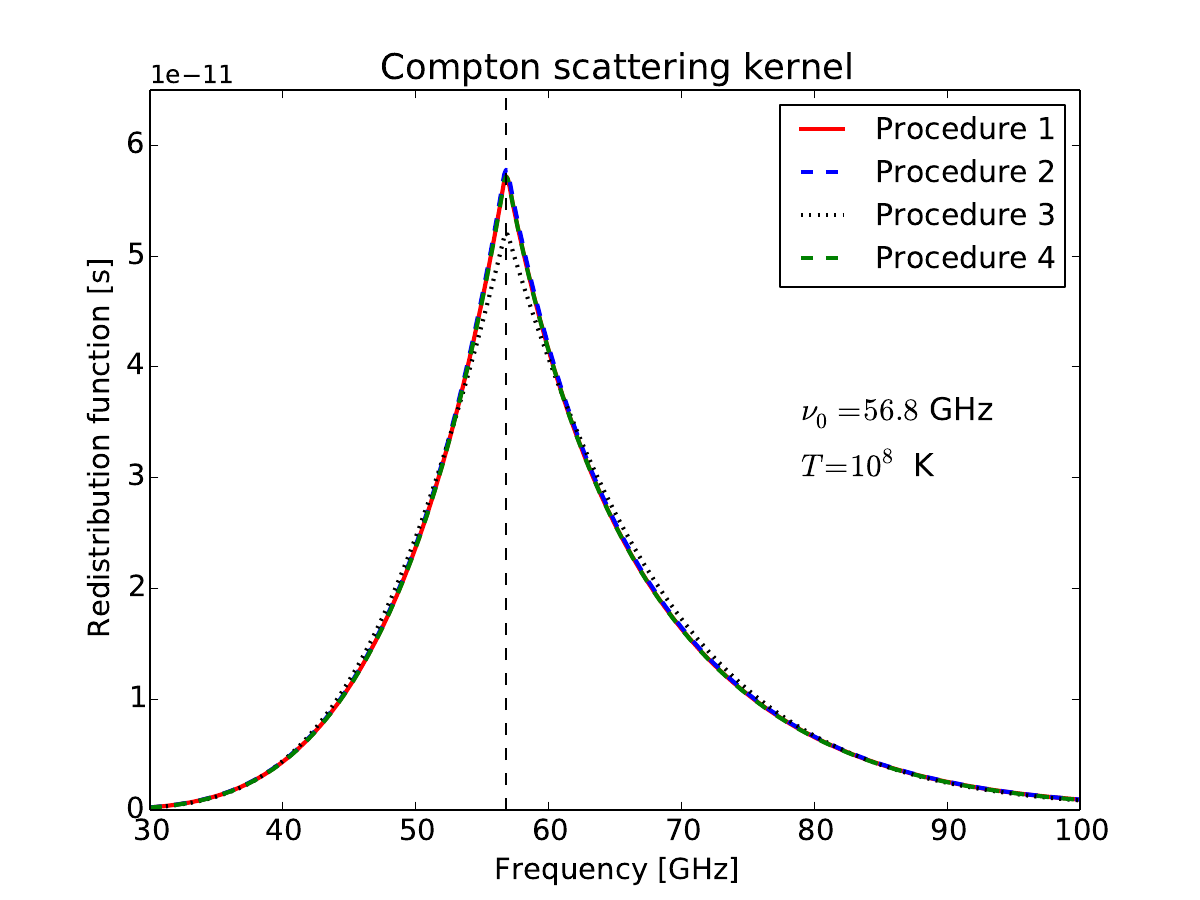}}}
    \vspace{0mm}
    \caption[]{Angle-integrated Compton scattering redistribution functions for microwave photons of
initial frequency $\nu = 56.8$ GHz, corresponding to a radiation temperature 2.768 K. Soft
photons are scattered here in the hot intra-cluster gas of electrons at temperature of $T=10^8$K. 
All the redistribution functions show the inverse Compton scattering effect. Again RF's by Guilbert 
(1981), Suleimanov et al. (2012), and Sazonov \& Sunyaev (2000) are practically identical, compare
the solid red line with the overlapping blue and green dashed lines.}
    \label{fig:cmb}
    \end{figure}

\section{Numerical results}

Figures 1-5 present runs of the angle-integrated redistribution functions 
$R_i (\nu,\nu^\prime)$ computed for a few sample gas temperatures and initial
photon energies. Note, that functions $R_i (\nu,\nu^\prime)$ are given always
in Hz$^{-1}$, while photon energies are either in keV or GHz (horizontal axis).

In all the figures, the redistribution function $R_2$ by Guilbert (1981) was drawn 
as a blue dashed line, while the exact function $R_1$ by Suleimanov et al. (2012) is 
represented by a red solid line. The shape and comparison of both redistribution 
functions for various assumed parameters is the most important part of this paper.
Curves showing the approximate functions $R_3$ and $R_4$ were indicated for
completeness (black dotted line and green dashed line, respectively).
 
Figs. 1-2 present the Compton redistribution functions for sample temperatures
$T= 1.8\times 10^7$ and $10^8$ K, which are typical for photospheres and
envelopes of hot X-ray bursting neutron stars. The initial photon energy is
similar to the energy of peak flux in the outgoing spectra (Fig. 1) or is a few
times higher (Fig. 2). Both functions $R_1$ and $R_2$ are practically identical
and overlap each other in the figures. Note, that in both cases reddening of
the scattered X-ray photons apparently dominates over the blue-shift.

Fig. 3 illustrates the Compton scattering of microwave photons of cosmic background
radiation (CMB) of the temperature $T=2.768$ K, scattered in hot gas in galaxy clusters
of $T=10^8$. Also here both functions $R_1$ and $R_2$ are identical. All the functions
$R_1 - R_4$ reproduce the inverse Compton effect and the blue-shift of microwave
photons dominates.

Fig. 4 demonstrates trace differences between both the essential redistribution
functions $R_1$ and $R_2$ for hard X-rays at temperature $T=10^8$ K, which corresponds 
to the deepest layers of hot neutron star atmospheres. More substantial differences
appear only at $T=10^9$ K or higher. As the example, Fig. 5 shows the Compton
redistribution functions for gamma ray photons of energy 1.24 MeV, significantly
exceeding the energy of the electron rest mass (511 keV).

  \begin{figure}
    \resizebox{0.7\hsize}{!}{\rotatebox{0}{\includegraphics{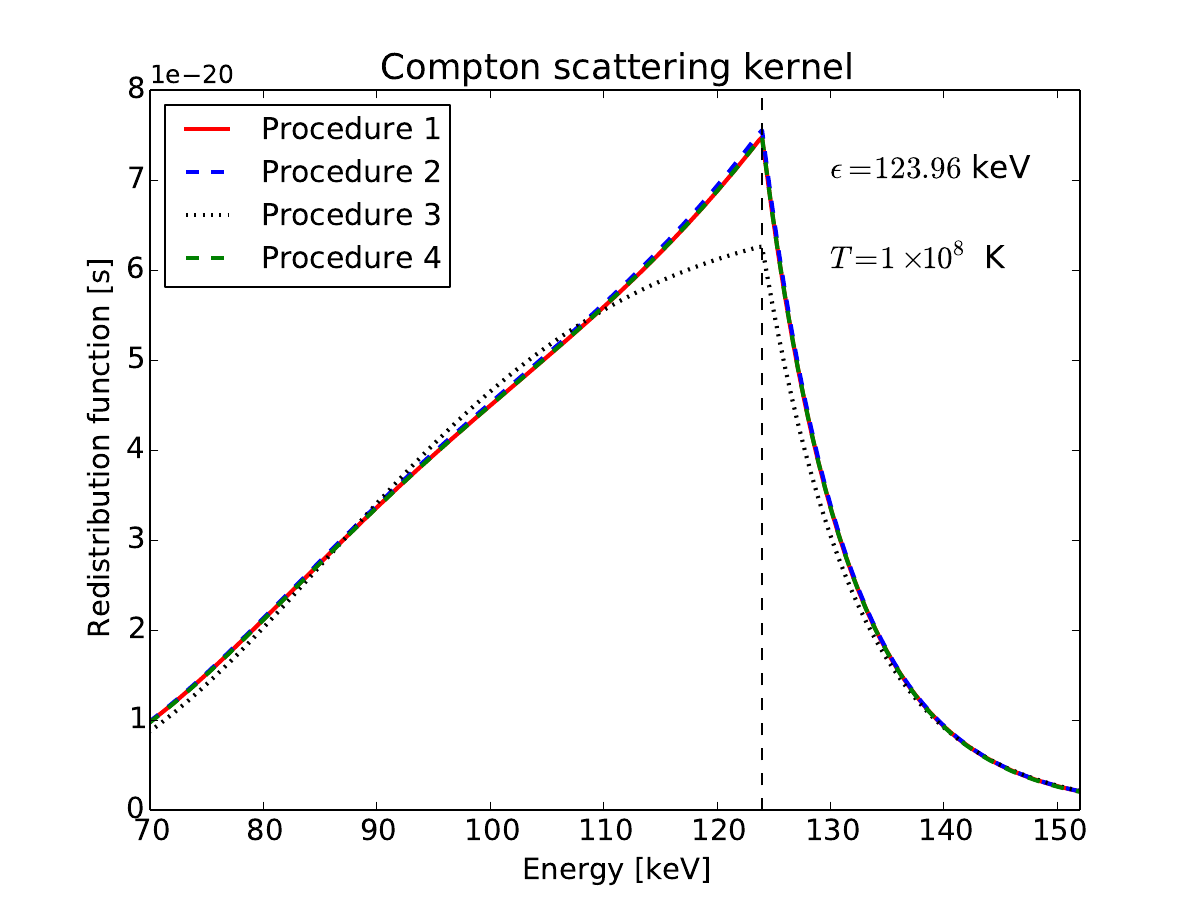}}}
    \vspace{0mm}
    \caption[]{Angle-integrated Compton scattering redistribution functions for X-ray photons of
initial wavelength $\lambda = 0.1 \AA$ (initial energy $\epsilon = 124.$ keV) in a gas of electron
temperature $T= 10^8$K. Formulae by Guilbert (1981) and Sazonov \& Sunyaev (2000) predict marginally 
different Compton redistribution function than the quantum mechanical formulae, compare the solid
red line and the dashed blue and dashed green lines. }
    \label{fig:fig4}
    \end{figure}
    
    \begin{figure}
    \resizebox{0.7\hsize}{!}{\rotatebox{0}{\includegraphics{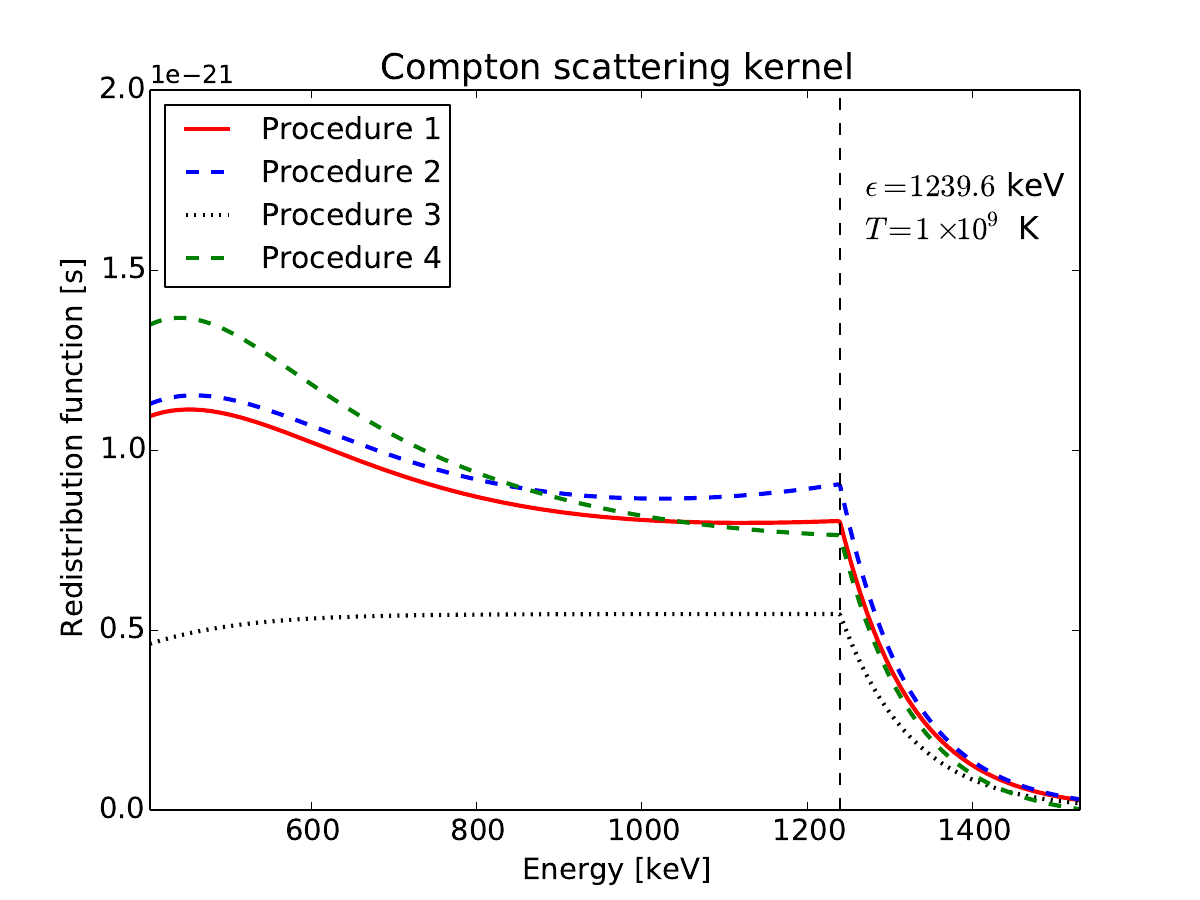}}}
    \vspace{0mm}
    \caption[]{Angle-integrated Compton scattering redistribution functions for X-ray photons
of initial wavelength $\lambda = 0.01 \AA$ (initial energy $\epsilon = 1.24$ MeV) in a gas
of electron temperature $T= 10^9$K. Only at such high $T$ do the exact Compton scattering 
redistribution function (solid red line) markedly differ from Guilbert's (1981) and 
Arutyunyan and Nikogosyan (1980) values. }
    \label{fig:fig5}
    \end{figure}

\subsection{Thermodynamic equilibrium}

Compton scattering redistribution function must obey the symmetry relation,
valid for electrons of maxwellian velocity distribution in thermodynamic
equilibrium (Pomraning 1973, Eqs. 8.1-8.2). The relation can be written as

\begin{equation}
\Delta (\epsilon,\epsilon_1,\eta,\Theta) =
  R_i (\epsilon,\epsilon_1,\eta) \epsilon^2 \exp(-\epsilon/\Theta ) -
  R_i (\epsilon_1,\epsilon,\eta) \epsilon_1^2 \exp(-\epsilon_1/\Theta) =0\, . 
\end{equation}

We numerically verified that equation for all Compton redistribution functions
$R_i$, $i=1,\ldots,4$ and computed tables of relative differences $\Delta_i /R_i$ for
all temperatures, initial energies $\epsilon$ and energy ranges $\epsilon^\prime$ 
shown in Figs. 1-5 and the cosine of scattering angles in the full range
[-1,+1]. The above identity was numerically reproduced here for $R_1$, $R_2$
and $R_3$ with the relative difference less than $10^{-19}$ (absolute value) 
almost everywhere in the parameter space, except at $\eta \rightarrow +1$,
where the relative difference could rise above $10^{-7}$.

Therefore, we conclude that the Guilbert's redistribution function $R_2$ 
described here fulfils the detailed balance condition. 

\section{Summary }

This paper presents four alternative formulae for calculating the photon 
redistribution function specific for the Compton scattering of unpolarized
light. Our considerations are valid in a perfect gas of electrons with isotropic
relativistic thermal velocities. These formulae were derived from published
papers on Compton scattering. 

The final scattering redistribution functions $R_i (\nu,\nu^\prime,\eta)$,
$i=1,\ldots,4$, are presented here in a unified dimensional form, which are ready 
to use in radiative transfer calculations ($i=1$ or 2). Approximate algorithms
No. 3-4 should not be used in accurate model atmosphere calculations. 

Furthermore, we present for the first time the correct set of equations
defining the Compton redistribution function ($R_2$) derived by Guilbert (1981).
The original paper was published with computational errors making his results
essentially useless. That method, now using correct equations, was applied in
the original Fortran code for model atmosphere computations of X-ray bursting
neutron stars (Madej 1991a,b; Madej et al. 2004).

We present also the exact quantum mechanical redistribution function $R_1$
(see Section \ref{sec:sec4}), defined in detail in Suleimanov et al. (2012).
We derived a new expression for $R_2$ by algebraic manipulation of the equations
given in their paper, which allowed us to perform numerical computations of $R_2$ in
a wide range of photon energies, from gamma rays down to radio waves.
Note, that our formulae are ideally suited for study of both hot stellar
atmospheres and spectral distortions of the cosmic microwave radiation
(Sunyaev-Zel'dovich effect), see Sazonov \& Sunyaev (1998), Chluba et al. (2012)
and Chluba \& Dai (2014). 

Some sample angle-integrated Compton scattering redistribution functions in hot plasma were
computed for gas temperatures $10^7 \le T \le 10^9$ K and initial photon
energies differing by many orders of magnitude. The resulting Figures 1-5
show that both algorithms by Guilbert (1981) and the exact quantum mechanical
equations produce the same redistribution functions, $R_1$ and $R_2$, for
Compton scattering in plasma at a temperature $T \le 10^8$ K. These are the
typical temperatures that occur in the atmospheres of X-ray bursters and
intracluster plasma. Only for higher temperatures, $T\ge 10^9$ K, do both
curves start to come apart. 

The Fortran 77 computer code for computations of all four Compton redistribution
functions, $R_1$ {\bf to $R_4$, } can be found at \rm{http://www.astrouw.pl/\~{}jm/software.html}.

\section*{Acknowledgments}

We are grateful to Dimitrios Psaltis, the referee, for helpful comments
and suggestions on our paper. We thank Sergey Sazonov for indication of
a fault in our preliminary figures and providing us results of his
calculations. 
This research was supported by Polish National Science Centre grants 
No. 2015/17/B/ST9/03422, 2015/18/M/ST9/00541 and  by Ministry of Science
and Higher Education grant W30/7.PR/2013. It received funding from the 
European Union Seventh Framework Programme (FP7/2007-2013) under 
grant agreement No.312789.

\section{Appendix A}

Guilbert (1981) defined the probability of scattering a photon of energy $\epsilon$ to 
energies between $\epsilon_1$ and $\epsilon_1 + d\epsilon_1$, from a direction ${\bf n}$ 
into a solid angle $d\Omega_1$, in a direction ${\bf n_1}$ along the raypath $dr$ is given by:
\begin{equation}
P\,d\epsilon_1 \, d\Omega_1 \, dr = \int \int (1-\beta \cdot {\bf n}) {{d \sigma}\over{d\Omega_1}}
M(\beta) {{\partial \beta} \over {\partial \epsilon_1}} d\epsilon_1 d\Omega_1 dr. 
\end{equation}

Energies are in units of the electron rest mass, $m_e c^2$; $\beta$ is the electron velocity in units of the speed of light; $M(\beta)$ is the electron velocity distribution and 
${d \sigma} / {d\Omega_1}$ is the differential cross-section for Compton scattering. 
\begin{equation}
(1-\beta \cdot {\bf n}) {{d \sigma}\over{d\Omega_1}} = {{3 \sigma_T} \over {16 \pi}}
{{\bar{X}} \over ({1-\beta \cdot {\bf n}})} { 1 \over \gamma^2 s^2}
\end{equation}
\begin{equation}
\bar{X} \equiv 2 -2 \left[ 1 - {\epsilon^2 \over 2s} (1-\cos{\theta}) \right]
{{1-\cos{\theta}} \over {\gamma^2 (1-\beta \cdot {\bf n})(1-\beta \cdot {\bf n_1})}} +
\left[ {{1-\cos{\theta}} \over {\gamma^2 (1-\beta \cdot {\bf n})(1-\beta \cdot {\bf n_1})}} \right]^2
\end{equation}
where 
\begin{equation}
s\equiv \epsilon/\epsilon_1;  \,\,\,\,\,\,\,\,  \cos{\theta}={\bf n} \cdot {\bf n_1}; \,\,\,\,\,\,\,\,  \gamma = {1 \over (1-\beta^2)^{1/2}} \nonumber 
\end{equation}
(see Babuel-Peyrissac \& Rouvillois 1969). 

For Compton scattering we have:
\begin{equation}
{1 \over s} = { {1-\beta \cdot {\bf n}} \over { \epsilon \gamma^{-1} (1-\cos{\theta}) + 1 - \beta \cdot {\bf n_1 }}},
\end{equation}
or, by rearranging terms




\begin{equation}
 \beta \cdot ({\bf n_1 } - s \, {\bf n}) = 1 - s
- \epsilon \gamma^{-1} (1-\cos{\theta}) 
\end{equation}

\vspace{2mm}

The above set of equations was transformed by Guilbert (1981) to 
\begin{equation}
P(\epsilon, s, \theta, x) = -{3\over{64\pi^2}}\,{1\over{Es^2}}\, 
  {x\over{K_2(x)}}  \int \limits_{\gamma_{\rm min}}^\infty 
  F(\epsilon, \epsilon_1,\theta,\gamma) \,\exp(-\gamma/\Theta)\, d\gamma \, , \\
\label{equ:Guilb}
\end{equation}
where the dimensionless gas temperature $x = m_e c^2 / {kT} $
Other auxiliary variables are defined as
\begin{eqnarray}
A &=& 1-s\, ,\hskip10mm B=\epsilon(1-\cos\theta)\, , \hskip10mm E = (1-2s\cos\theta+s^2)\, ^{1/2}\, ,\\
\gamma_{\rm min} &=& \left\{ \left[ {E^2 \over{E^2 + B^2}} \left( 1-
  {A^2 \over {E^2 +B^2}} \right) \right]^{1/2} - {AB \over{E^2 +B^2}} \right\}^{-1} \, .
\label{equ:int2}
\end{eqnarray}
Relativistic Compton redistribution function then equals to
\begin{equation}
F(\epsilon, \epsilon_1,\theta,\gamma) = (1-\beta_z)
  \int \limits_0^{2\pi} {{\bar{X}} \over ({1-\beta \cdot {\bf n}})} \, d\phi \, ,
\hspace{15mm} {\rm where} \>\>\> \beta_z = (A+B\gamma^{-1})/E\, .
\end{equation}

\vspace{2mm}

Function $F(\epsilon,s,\theta,\gamma)$ can be 
expressed by the trinomial (Guilbert 1981)
\begin{equation}
F(\epsilon,s,\theta,\gamma)=2a I_1 + {2a \over \gamma^2} (1- \cos \theta) 
\left[1 - {\epsilon^2 \over 2s} (1- \cos \theta) \right] I_2 + 
{a \over \gamma^4} (1 - \cos \theta)^2 I_3 ,
\end{equation}

\noindent Let us define variables
\begin{eqnarray}
 a&=& 1-(\cos\theta-s)\,{1-s+\epsilon\, (1-\cos \theta)/\gamma \,\over {1-2s\cos\theta +s^2}} \\
 b&=& -\left( {\gamma^2-1\over\gamma^2}-{(1-s+\epsilon (1-\cos\theta)/\gamma)^2
    \over{1-2s\cos\theta+s^2}} \right)^{1/2}
    \left( {1-\cos^2 \theta \over{1-2s\cos\theta+s^2}} \right)^{1/2}  \\
 c&=& 1 -(1-s\cos\theta)\,{1-s+\epsilon\, (1-\cos \theta)/\gamma \,\over{1-2s\cos\theta+s^2}} \\
 d&=& b\,s   \\ 
 y&=& { b \over \sqrt{a^2-b^2}} + {d \over \sqrt{c^2-d^2}} 
\end{eqnarray}

\noindent then the partial integrals
\begin{eqnarray}
I_1 & = & { {2 \pi} \over \sqrt{a^2-b^2}}   \\
I_2 & = & {{ 2 \pi} \over { y (a^2-b^2)^2(c^2-d^2)}}  \left[ a(bc + ad) + b(ac + bd) 
- {{ab(bc+ad)} \over {y \sqrt{a^2-b^2}}}  \right]      \\
I_3 & =& 
{{\pi} \over { y ( a^2-b^2)^3 (c^2-d^2)^2}} ~\times ~ \left( ~  
\left[ a(bc+ad) + b(ac+bd) - {{ ab(bc+ad)} \over {y \sqrt{a^2-b^2}}} \right] \right. \\
& \times &
\left. \left[ {{ 2abcd} \over {y^2 \sqrt{a^2-b^2}\sqrt{c^2-d^2}}} - {{4 a c d 
}\over {y
\sqrt{c^2-d^2}}} 
- {{2 a b c } \over {y \sqrt{a^2 - b^2}}} + 8 a c \right] \right. \nonumber \\  
&+ &  
\left. \left[ 2 a d + 2 b c - { {b^2 c + 2 a b d} \over {y \sqrt{a^2 - b^2}}}+ 
{ {a^2 b (b c + a d)} \over {y (a^2 -b^2)^{3/2} }} -
{ { a^2 b^2 (b c + a d )} \over { y^2 (a^2-b^2)^2 }} \right]  
   \times (a^2-b^2) c \left( {d \over {y \sqrt{c^2-d^2}}}  -2 \right) \right. \nonumber \\
  & + &
 \left.   \left[ 2 a b   - {{ab^2} \over {y \sqrt{a^2-b^2}}} - 
  {{abcd(bc+ad)} \over {y^2 \sqrt{a^2-b^2} (c^2-d^2)^{3/2}}} \right] \times \left[
 {{ab(c^2-d^2)} \over {y \sqrt{a^2-b^2}}}-4a(c^2-d^2) \right]
 + 2b (a^2-b^2) (c^2-d^2)  \right.  \nonumber \\
 & - & 
 \left. {b^2 \over y} \sqrt{a^2-b^2} (c^2-d^2) - a^2 b^3 { {(c^2-d^2)} \over {y^2 (a^2-b^2)}} 
 + a^2 b^2 {{(c^2-d^2)}\over{y \sqrt{a^2-b^2}}} -{ {bcd 
(bc+2ad)}\sqrt{a^2-b^2} \over{y^2
\sqrt{c^2-d^2}}}\right. \nonumber \\
 & -& 
 \left.  {{2a^2b^2 cd (bc+ad)} \over {y^3 (a^2 -b^2) \sqrt{c^2-d^2}}} +
 {{a^2bcd (bc+ad)} \over {y^2 \sqrt{a^2-b^2} \sqrt{c^2-d^2}}} \right) \nonumber
\label{equ:i3}
\end{eqnarray}

The above algorithm was defined by Guilbert (1981), who in his paper presented
an erroneous expression for the integral $I_3$. Herein, we correctly define the
coefficient $I_3$, which was used in a series of papers on model atmospheres of
bursting neutron stars (Madej 1989, 1991a, 1991b).

\bsp	
\label{lastpage}
\end{document}